\documentclass[conference]{IEEEtran}
\usepackage{balance}
\newcommand{\Var}{{\textsc{Var}}}
\newcommand{\Form}{{\textsc{Form}}}

\newcommand{\ie}{\textit{i.e.}}
\usepackage{mathrsfs} 

\usepackage[cmex10]{amsmath}
\usepackage{amssymb}
\usepackage{eurosym}
\usepackage{color}
\usepackage{graphicx}

\begin{document}
%
\title{Querying with {\L}ukasiewicz logic}

\author{\IEEEauthorblockN{Stefano Aguzzoli\\and Pietro Codara}
\IEEEauthorblockA{Dipartimento di Informatica\\
Universit\`{a} degli Studi di Milano, Italy\\
\{aguzzoli, codara\}@di.unimi.it}
\\
\IEEEauthorblockN{Diego Valota}
\IEEEauthorblockA{Artificial Intelligence Research Institute (IIIA)\\ CSIC, Spain\\
diego@iiia.csic.es}
\and
\IEEEauthorblockN{Tommaso Flaminio\\and Brunella Gerla}
\IEEEauthorblockA{Dipartimento di Scienze Teoriche e Applicate\\
Universit\`{a} dell'Insubria,\\
Varese, Italy\\
\{tommaso.flaminio, brunella.gerla\}@uninsubria.it}
}


%


\maketitle

\begin{abstract}
In this paper we present, by way of case studies, a proof of concept,
based on a prototype working on a automotive data set, aimed at showing the potential usefulness
of using formulas of {\L}ukasiewicz propositional logic to query
databases in a fuzzy way.
Our approach distinguishes itself for its stress on the purely linguistic,
contraposed with numeric, formulations of queries. Our queries are
expressed in the pure language of logic, and when we use (integer) numbers,
these stand for shortenings of formulas on the syntactic level,
and serve as linguistic hedges on the semantic one.
Our case-study queries aim first at showing that each numeric-threshold
fuzzy query is simulated by a {\L}ukasiewicz formula. Then they focus
on the expressing power of {\L}ukasiewicz logic which easily allows
for updating queries by clauses and for modifying them through a potentially
infinite variety of linguistic hedges implemented with a uniform syntactic
mechanism. Finally we shall hint how, already at propositional level,
{\L}ukasiewicz natural semantics enjoys a degree of reflection, allowing
to write syntactically simple queries that semantically work as meta-queries
weighing the contribution of simpler ones.
\end{abstract}

%
\IEEEpeerreviewmaketitle

\section{Introduction, and motivation}

The aim of this paper is to give a rather informal
presentation of a natural semantics for
{\L}ukasiewicz logic (contraposed with the formal [0,1]-valued
semantics) where formulas are interpreted as fuzzy queries
to a database.

In this framework,
database entries (the rows of a table) are identified
with truth-value assignments, or possible worlds,
and the evaluation of a query is simply the
truth-value of the formula encoding the query
in the considered possible worlds.

The negation connective plays the r\^{o}le
of asking for the opposite quality to the one
being negated.
Lattice connectives behave much like their
Boolean counterparts, creating unions and intersections
of answer sets.
Monoidal, non-idempotent connectives, which characterise
{\L}ukasiewicz logic, are used both to implement
a mechanism to formulate an infinite variety of linguistic hedges,
and to act as a comparison and weighing operator between
simpler queries.

We apply these notions to a prototype system that
allows to query a database of cars via formul\ae{}
of pure propositional {\L}ukasiewicz logic.
We analyse some examples to show how users
can benefit from the flexibility and expressive power
of this language.

We remark that, at least in principle, our queries
are purely linguistic objects, translatable in
natural language, where the linguistic hedges
are translated as applications (maybe iterated)
of {\em somewhat} and {\em very},
while the monoidal connective $\ominus$
corresponds to a query asking for objects (cars in our example application)
that satisfy some given properties {\em much more} than other ones.
We strike a comparison of our purely linguistic queries with
numeric-threshold based queries, and show how the latter are
very easily replicated in our chosen language.

The paper is organised as follows.
Section \ref{sec:luklogic} is a brief introduction
to infinite-valued propositional {\L}ukasiewicz logic,
with statement of the most relevant results which
are at the base of our proposed natural semantics for
{\L}ukasiewicz logic based queries.

Section \ref{sec:db} shortly describes our
prototype implementation to query a database
of cars.

Section \ref{sec:cases} analyses several
queries to illustrate why we think
that querying a database with formulas of {\L}ukasiewicz
may be useful, and reflect the proposed natural semantics.

\section{{\L}ukasiewicz logic in brief}\label{sec:luklogic}

\emph{{\L}ukasiewicz \textup{(}infinite-valued propositional\textup{)} logic} is a non-classical many-valued system going back to the 1920's, cf.\ the early survey  \cite[\S 3]{luk}, and its annotated  English translation  in \cite[pp.\ 38--59]{tar}.
 The standard modern reference for  {\L}ukasiewicz logic is \cite{mun1}, while  \cite{mun2} deals
with topics at the frontier of current research. {\L}ukasiewicz logic can also be regarded as a member of a larger hierarchy of many-valued logics that was systematised by Petr H\'{a}jek in the late Nineties, cf.\ \cite{hajek}, and later extended by Esteva and Godo in \cite{estevagodo}; see also \cite{HB1,HB2}. Let us recall some basic notions.

\smallskip Let us fix once and for all the countably infinite set of propositional variables:
\[
\Var = \{X_1,X_2,\ldots,X_n,\ldots\}\,.
\]
Let us  write $\bot$ for the logical constant {\it falsum}, $\neg$ for the unary negation connective, and $\to$ for the binary implication connective. The set $\Form$ of (well-formed) formul\ae{}
is defined exactly as in classical logic over the language $\{\bot,\neg,\to \}$.
Derived connectives $\top, \vee, \wedge, \leftrightarrow, \oplus, \odot, \ominus$ are defined in the following table, for every formula $\alpha$ and $\beta$:
\begin{table}[htf]
\begin{center}\begin{tabular}{|c|c|c|}
\hline {\bf Derived connective} &  {\bf Definition} \\
\hline
\hline
$\top$  & $\neg \bot$\\
\hline
$\alpha\vee \beta$   & $(\alpha \to \beta)\to \beta$\\
\hline
$\alpha\wedge \beta$   &$ \neg (\neg \alpha \vee \neg \beta)$\\
\hline
$\alpha\leftrightarrow \beta$ &$(\alpha \to \beta ) \wedge (\beta \to \alpha)$\\
\hline $\alpha\oplus \beta$  & $\neg \alpha \to \beta$ \\
\hline $\alpha\odot \beta$   & $\neg (\neg \alpha \oplus \neg \beta)$\\
\hline $\alpha\ominus \beta$    & $\neg (\alpha\to\beta)$ \\
\hline \end{tabular} \smallskip
\end{center}
\caption{ Derived connectives in {\L}ukasiewicz logic.}\label{table:01connectives}

\end{table}

\smallskip
Let us present the $[0,1]$-valued semantics of \L{}ukasiewicz logic. An \emph{atomic assignment},
or \emph{atomic evaluation}, is an arbitrary function $\overline{w}\colon \Var \to [0,1]$.  Such an atomic evaluation is uniquely extended to an \emph{evaluation} of all formul\ae, or \emph{possible world}, \ie\ to
a function $w\colon \Form \to [0,1]$ \footnote{Since we are working with a purely truth-functional propositional logic we safely consider {\em evaluation} and {\em possible world} as synonymous.},
via the compositional rules:
\begin{align*}
w(\bot)&=0\,,  \\
w(\alpha\to\beta)&=\min{\{1,1-(w(\alpha)-w(\beta))\}}\,,\\
w(\neg\alpha)&=1-w(\alpha)\,.
\end{align*}
It follows by trivial computations that  the
formal semantics of derived connectives is the one reported in Table \ref{table:01connectives}.
\emph{Tautologies} are defined as those formul\ae\ that evaluate to $1$ under every evaluation.
\begin{table}[htf]
\begin{center}\begin{tabular}{|c|c|c|}
\hline {\bf Notation}    & {\bf Formal semantics} \\
\hline\hline
$\bot$& $w(\bot)=0$\\
\hline
$\top$  & $w(\top)=1$\\
\hline
$\neg \alpha$   & $w(\neg\alpha)=1-w(\alpha)$\\
\hline$\alpha\to \beta$   & $w(\alpha\to\beta)=\min{\{1,1-(w(\alpha)-w(\beta))\}}$\\
\hline
$\alpha\vee \beta$   & $w(\alpha\vee \beta)=\max{\{w(\alpha),w(\beta)\}}$\\
\hline
$\alpha\wedge \beta$   &$w(\alpha\wedge \beta)=\min{\{w(\alpha),w(\beta)\}}$\\
\hline
$\alpha\leftrightarrow \beta$ &$w(\alpha\leftrightarrow\beta)=1-|w(\alpha)-w(\beta)|$\\
\hline $\alpha\oplus \beta$  & $w(\alpha\oplus\beta)= \min{\{1,w(\alpha)+w(\beta)\}}$ \\
\hline $\alpha\odot \beta$   & $w(\alpha\odot\beta)= \max{\{0,w(\alpha)+w(\beta)-1\}}$\\
\hline $\alpha\ominus \beta$    & $w(\alpha\ominus\beta)= \max{\{0,w(\alpha)-w(\beta)\}}$ \\
\hline \end{tabular} \smallskip
\end{center}
\caption{ Formal semantics of connectives in {\L}ukasiewicz logic.}\label{table:01connectives}

\end{table}
%

Each propositional formula $\varphi$ whose occurring variables are in $\{X_1,X_2,\ldots,X_n\}$
uniquely determines a term function
$$\bar{\varphi} \colon [0,1]^n \to [0,1]\,,$$
by the following inductive prescription, for every $(t_1,\ldots,t_n) \in [0,1]^n$.
\begin{enumerate}
\item If $\varphi = X_i$ then $\bar{\varphi}(t_1,\ldots,t_n) = t_i$.
\item If $\varphi = \bot$ then $\bar{\varphi}(t_1,\ldots,t_n) = 0$.
\item If $\varphi = \neg\psi$ then $\bar{\varphi}(t_1,\ldots,t_n) = 1 - \bar{\psi}(t_1,\ldots,t_n)$.
\item If $\varphi = \psi \to \vartheta$ then $\bar{\varphi}(t_1,\ldots,t_n) = \min\{1,1 -(\psi(t_1,\ldots,t_n) - \vartheta(t_1,\ldots,t_n))\}$.
\end{enumerate}
This definition implies that $w(\varphi) = \bar{\varphi}(\bar{w}(X_1),\ldots,\bar{w}(X_n))$ for all possible worlds $w$.

McNaughton's Representation Theorem \cite{mcnaughton} states that the class of $n$-variable term functions
$\bar{\varphi} \colon [0,1]^n \to [0,1]$ coincides with the class of all functions
$f \colon [0,1]^n \to [0,1]$ that are continuous (in the standard Euclidean topology), and piecewise linear with integer coefficients,
that is, there exist finitely many linear polynomials $p_1,p_2,\ldots,p_u \colon [0,1]^n \to [0,1]$  such that
each $p_i$ has the form $p_i(t_1,\ldots,t_n) = b_i + \sum_{j=1}^n a_{i,j} t_i$ for all coefficients $a_{i,j}$ and $b_i$ being integers,
and a function $\iota \colon [0,1]^n \to \{1,2,\ldots,u\}$ such that
$$f(t_1,\ldots,t_n) = p_{\iota(t_1,\ldots,t_n)}(t_1,\ldots,t_n)\,,$$
for all $(t_1,\ldots,t_n) \in [0,1]^n$.

Menu-Pavelka's Theorem \cite{pavelka} states that {\L}ukasiewicz logic is
characterised in the H\'{a}jek's hierarchy BL of Basic Fuzzy Logics, or in the even larger
Esteva and Godo's hierarchy MTL \cite{estevagodo} of Monoidal $t$-norm-based Logics,
as the unique logic having continuous term functions.
This fact, together with involutiveness of negation,
that is
$$w(\neg \neg \varphi) = w(\varphi)$$
for all possible worlds $w$,
and the simultaneous failure of idempotency for the monoidal connectives
$\oplus$ and $\odot$,
not to mention the deep connection with lattice-ordered abelian groups \cite{mun1},
which allows to model real arithmetic on the unit interval,
renders {\L}ukasiewicz logic a very interesting tool to implement
fuzzy-based applications.

In this paper we shall focus on some other properties of {\L}ukasiewicz
logic that further support the notion that this logic may constitute
the ideal theoretical backbone to certain fuzzy-based applications.

Our first concern is actually rather philosophical and constitutes in
our opinion a defensible rebuttal against the frequent attack to fuzzyness
consisting in the observation that graded, or $[0,1]$ fuzzy truth-values
are arbitrary or have no meaning at all.
We counter this statement considering maximally consistent theories.

A theory $\Theta$ in {\L}ukasiewicz logic is a deductively closed set of
formulas, that is, if a formula $\varphi$ is such that
$w(\varphi) = 1$ for all possible worlds $w$ such that $w(\vartheta) = 1$
for all $\vartheta \in \Theta$, then $\varphi \in \Theta$, too.

A theory $\Theta$ is maximally consistent if it cannot be enlarged
without losing consistency, that is, $w(\varphi) < 1$ for
all $\varphi \not\in \Theta$ and
for all possible worlds $w$ such that $w(\vartheta) = 1$.

A set of postulates  for a theory $\Theta$ is
a set of formulas $\Gamma \subseteq \Theta $ such that
$\Theta$ is the deductive closure of $\Gamma$,
that is, $\Theta$ contains exactly all formulas
$\vartheta$ such that $w(\vartheta) =1$ for
all possible worlds $w$ for which $w(\gamma) = 1$
for all $\gamma \in \Gamma$.

In \cite{marra} Marra points out that the set of
maximally consistent theories written in the variables $\{X_1,X_2,\ldots,X_n\}$
corresponds bijectively to the set of all $n$-tuples $[0,1]^n$.
Moreover, each $\bar{t} = (t_1,\ldots,t_n) \in ([0,1] \cap \mathbb{Q})^n$
corresponds to a theory $\Theta_{\bar{t}}$ having a set of postulates
$\Gamma_{\bar{t}} = \{\gamma_{\bar{t}}\}$ for a suitable formula $\gamma_{\bar{t}}$.

When we consider formulas in just one variable, the above bijection
tells us that each truth-value $\delta \in [0,1]$ corresponds,
in the formal semantics of {\L}ukasiewicz logic, exactly
to one maximally consistent theory. That is, the choice
of a value in $[0,1]$ is canonically and consistently
reflected in the choice of a maximally consistent theory.
So, any truth value has a canonical and fixed semantics,
formed by the formulas in the corresponding theory.
{\em Vice versa}, each truth-value is described linguistically
by a set of formulas.
We refer the reader to \cite{marra} for a thorough treatment
of this topic. In this paper we are specially interested
in some viable pragmatic consequences of this correspondence.

As a matter of fact, for each rational $\delta \in [0,1] \cap \mathbb{Q}$,
the formula $\gamma_{\bar{\delta}}$ can be constructed
as $\alpha \vee \beta$, where $\alpha$ and $\beta$ are built
from $X_1$ and $\neg X_1$ using only the {\em minus} connective $\ominus$.

As we shall see, the $\ominus$ connective will play a major r\^{o}le
in the semantics we mean to use to query databases with
{\L}ukasiewicz logic.

\subsection{An intended semantics for {\L}ukasiewicz logic}\label{sec:intsem}

We shall sketch in this section a natural semantics, which corresponds
and interprets the formal semantics given in the previous section,
and prepare the way to use it to express and interpret fuzzy queries
to a database.

\subsubsection{Possible Worlds, Queries and Answer Sets}

Possible worlds $w \colon \Form \to [0,1]$ corresponds bijectively
to atomic assignments $\bar{w} \colon \Var \to [0,1]$.
Clearly the value $w(\varphi)$ depends only on the finitely many variables occurring in $\varphi$.
We shall identify atomic assignments, and hence, possible worlds, with the fuzzy entries in the rows of a database table.
More precisely, if a row $r$ belongs to a table where all the columns with values ranging in $[0,1]$ are $c_1,c_2,\ldots,c_u$,
then we consider $r \colon \{c_1,c_2,\ldots,c_u\} \to [0,1]$ as the atomic assignment corresponding to a possible world
where to evaluate our queries, which will be formulas over the variables $\{c_1,c_2,\ldots,c_u\}$.
It is then straightforward that the answer-set to a query $\varphi$ over some table
is formed by the rows $r$ such that $r(\varphi) = 1$.
Notice that in the next section for sake of simplicity
we shall speak of answer set also when referring
to (the top part of) the set of rows $r$ ranked
by the values $r(\varphi)$.

\subsubsection{Variables, and the Negation Connective}

But, what entries in the database table shall we consider as fuzzy?
In our approach, we shall consider a column to be fuzzy valued if it
corresponds to a graded (and $[0,1]$-normalised) property which has an opposite.
For instance, {\em tall} has as its opposite {\em short}, with
their obvious general meanings, while {\em red},
the property of being red, might not have a natural opposite
(clearly, this depends on context, and we may always setup
for an artificial opposite {\em non-red}).
Of each pair of opposites $(p,q)$ we clearly only need to store
one value $r(p)$ for each row $r$, stating how much $r$ is $p$.
Clearly the degree of $r$ being $q$ shall be given by $\neg r(p)$.

\subsubsection{Conjunction and Disjunction}

One can form conjunctions and disjunctions of simpler queries just by using
the lattice connectives. Recall that $r(\varphi \vee \psi) = \max\{r(\varphi),r(\psi)\}$
and $r(\varphi \wedge \psi) = \min\{r(\varphi),r(\psi)\}$.
Observe that the resulting answer sets correspond, as in the crisp Boolean case,
to unions and intersections of the simpler answer sets.

\subsubsection{Basic Literals and Linguistic Hedges}\label{sec:literals}

One of the most interesting features of fuzzy querying is the capability
of using linguistic hedges such as {\em somewhat} or {\em very}.
{\L}ukasiewicz logic offers a uniform and syntactically simple
mechanisms to implement a collection of infinitely many
such hedges, through the use of basic literals.

For each integer $k > 0$ and each formula $\varphi$ let $1 \varphi = \varphi$
and $(k+1) \varphi = \varphi \oplus k\varphi$. Analogously,
let $\varphi^1 = \varphi$ and $\varphi^{k+1} = \varphi \odot \varphi^k$.
Basic literals \cite{literals} are the class of formulas described inductively by the following.
\begin{enumerate}
\item Each variable $X_i$ is a basic literal.
\item If $\varphi$ is a basic literal then $k \varphi$ is a basic literal.
\item If $\varphi$ is a basic literal then $\varphi^k$ is a basic literal.
\end{enumerate}

Notice that $w(2 \varphi) = 1$ iff $w(\varphi) \geq 1/2$, whence
we use $2 \varphi$ to model {\em somewhat $\varphi$}.
Analogously, $w(\varphi^2) = 0$ iff $w(\varphi) \leq 1/2$,
and we use it to model {\em very $\varphi$}.
Moreover, we can intensify our notion of being {\em somewhat $\varphi$},
by using $k \varphi$ for some $k > 2$.
Analogous intensifications of {\em very $\varphi$} are possible,
and moreover we can create more complex hedges by means
of general basic literals. For instance
$2(\varphi^2)$ may model being somewhat but not extremely $\varphi$
(or, {\em somewhat very $\varphi$}, to play the game just syntactically).

Clearly, complex basic literals are very hard to translate
in comprehensible linguistic expressions, but this is true
also of natural language sentences built with many
occurrences of words such as {\em somewhat} and {\em very}.

Being linguistic hedges, basic literals can be used
to express threshold queries, and, as a matter
of fact they can reproduce the answer set of
any numeric-threshold query.
Consider any such query $r(\varphi) \geq  \delta$ for some $\delta \in [0,1]$
(clearly, in applications, $\delta$ is rational, but the theory of basic literals
deals with irrational $\delta$, too).
As the database table contains finitely many rows,
either the answer set does not discard any row, or there is a row $r_1$
such that
$r_1(\varphi) = \max\{r(\varphi) < \delta\}$. In this case,
pick any two rationals $q_1,q_2$ with $r_1(\varphi) \leq q_1 < q_2 \leq \delta$.
As proved in \cite{literals} there is a basic literal $\psi$
such that $r(\psi(\varphi)) = 1$ iff $r(\varphi) \geq q_2$,
and $r(\psi(\varphi)) = 0$ iff $r(\varphi) \leq q_1$.
This shows that any query $r(\varphi) \geq \delta$ can be reproduced.
Analogously, any query $r(\varphi) \leq \delta$  can be reproduced, too.

\subsubsection{The $\ominus$ Connective and Comparisons}

Usually, the algebraic treatment of {\L}ukasiewicz logic
is conducted electing one of the non-idempotent operations of the formal semantics
as primitive. Tipically the chosen one is $\oplus$, as in MV-algebras \cite{mun1},
or $\odot$ as in involutive BL-algebras \cite{hajek}, or $\to$ as in Wajsberg hoops
\cite{mun1}. We base our semantics on the $\ominus$ connective instead,
for it affords us to model comparisons between queries.

Recall that for any possible world (or row of a database table) $r$,
we have $r(\varphi \ominus \psi) = \max\{0,r(\varphi) - r(\psi)\}$,
that is, the semantics of $\ominus$ is truncated difference.

The truncated difference $r(\varphi \ominus \psi)$ evaluates to $1$ iff
$r(\varphi) = 1$ and $r(\psi) = 0$, and goes linearly to $0$,
which is achieved when $r(\varphi) = r(\psi)$.
It is then very tempting to read $r(\varphi \ominus \psi)$
as
$$r \mbox{ is } \varphi \mbox{ much more than it is } \psi\,,$$
or
$$\mbox{ in the world } r,\  \varphi \mbox{ holds much more than } \psi\,.$$
Notice the use of {\em much} since $r$ will appear in the answer set
of $\varphi \ominus \psi$ iff $r(\varphi) = 1$ and $r(\psi) = 0$.
This causes no real problem, since we can get a milder version attenuating
the impact of the hedge {\em much} by use of the hedge {\em somewhat}.

We can then use $\ominus$ to model queries that gauge the
difference between the values of other queries.
We recall here that the linguistic power of $\ominus$
afford a genuine interpretation of fuzzy truth-values,
since, as pointed out in the previous section,
each rational in $[0,1]$ encodes the meaning
of a formula built from $X_1$ and $\neg X_1$
by means of a disjunction of two
subformulas each one of them built only
using, possibly iterated, occurences of $\ominus$.
Each irrational is encoded
by infinitely many formulas built in a similar
fashion.

\section{A system to query a database with formulas of {\L}ukasiewicz logic}\label{sec:db}

To empirically test the intended semantics of {\L}ukasiewicz logic given in Section~\ref{sec:intsem},
we have implemented a simple web interface that translates {\L}ukasiewicz logic formulas
into SQL statements.
This web application has been developed following the standard programming pattern \emph{Model-View-Controller} (MVC) \cite{MVC}.
To facilitate the development we have used the PHP Phalcon\footnote{Phalcon is a PHP module developed in C:
this makes Phalcon one of the fastest PHP web frameworks.}
programming framework on the server side, hence all the \emph{controllers} in our programming paradigm are written in PHP language.
On the other hand, on the client side of the application HTML and Javascript are the languages of election. 
To make \emph{views} more user-friendly we have employed different libraries, such as JQuery,
Bootstrap, MathJax and MathLex.

Finally, our \emph{model} is a single database table where we have collected cars data.
The table fields with associated datatypes are summarised in the first two columns of Table~\ref{tab:cardata}.
The database table contains 4684 records. We use MySQL as DBMS.

\begin{table}
\caption{
}\label{tab:cardata}
\begin{tabular}{llr}
\hline
 Field  &  Type &   Associated Variable  \\
\hline
 id  & int(10) unsigned & --\\
 manufacturer &                varchar(50)      & --\\
 model	&	         varchar(50)     & --\\
 trim    &       varchar(200)    & --\\
 price    &             int(11)   &  X0\\
 length    &          int(11)      &  X1\\
 width      &        int(11)        & X2\\
 height      &          int(11)      &  X3\\
 fuel tank     &          int(11)      &  X4\\
 seating capacity         &         tinyint(4)    & X5\\
 car segment    &           varchar(50) &  --\\
 drive           &    varchar(50)     & --\\
 fuel      &    varchar(50)     & --\\
 cubic capacity - cc       &      int(11)       &  X6\\
 horsepower         &       int(11)      &  X7\\
 power               & int(11)         & X8\\
 environmental classification        &   varchar(10)   & --\\
 co2 emission &         int(11)         &X9\\
 gearbox         &        varchar(50)     &--\\
 max speed      &     smallint(6)     &X10\\
 acceleration 0/100  &  decimal(5,2)   & X11\\
 urban cycle consumption    &     decimal(5,2)   &X12\\
 extra-urban cycle consumption     &     decimal(5,2)   &X13\\
 combined cycle consumption        &  decimal(5,2)    & X14
\end{tabular}
\end{table}

The web interface has two key pages: one allows to normalise the numerical fields of the table,
and the second allows to write a logical formula and submit it as a query to the database.

\begin{figure}[h!]
 \centering
 \includegraphics[scale=0.28]{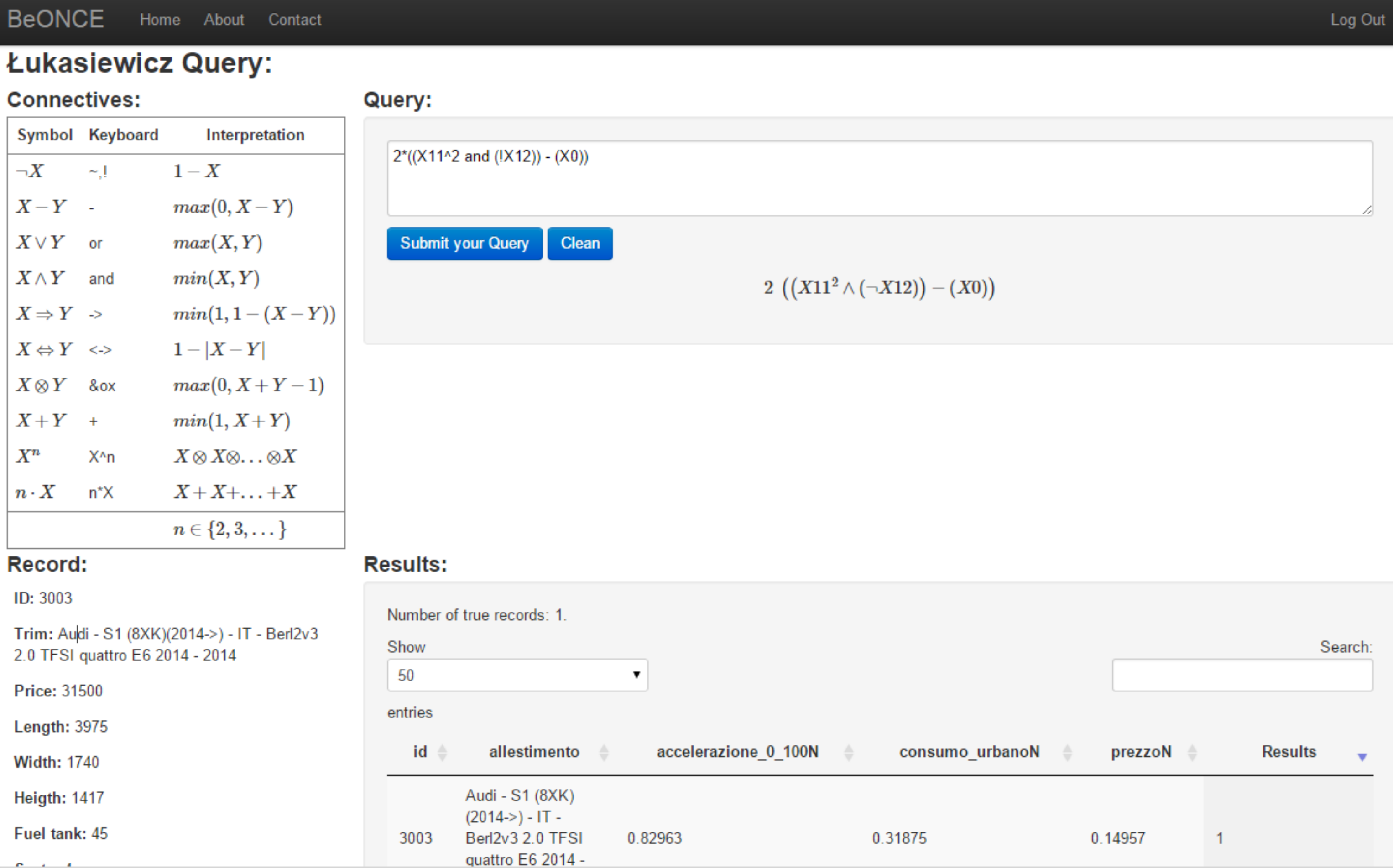}
 \caption{The web page to submit queries}
\end{figure}

The first page shows the maximum $M$ and the minimum $m$ values of each
numerical field stored in the table, allowing the user to choose his own maximum
$M_u \leq M$ and minimum $m_u \geq m$ to normalise every field according to his personal bias.
For instance, the field \emph{max speed} has a maximum value $M$ equal to $350$.
One can choose to fix $M_u=250$; in this way all cars with max speed $\geq 250$
will be considered as the fastest cars stored in the database (they are fast in degree $1$).
The normalised values $n_i$ are obtained by the standard linear formula
$n_i = (v_i - m_u)/ (M_u - m_u)$, where $v_i$ is the value of the considered field
at the $i$-th row.
There are some numerical fields where the best result is not given by the
maximum value but by the minimum (for istance \emph{acceleration 0/100}).
In such cases, by ticking a checkbox the user can reverse the normalisation of the given field.
That is, $n_i = 1 - ((v_i - m_u)/ (M_u - m_u))$.

The second page is the one where the database can be queried through the use of logical formulas.
Therefore, in the page appears a list of variables \texttt{XN}, with \texttt{N}$\in\{0,\dots,14\}$.
Every variable \texttt{XN} represents a precise field of the table.
The third column of Table~\ref{tab:cardata} gives the association between variables and fields.
Then, the user is allowed to write a logical formula using such variables and the logical connectives
described in Section~\ref{sec:luklogic}.
The correspondence between connectives and \emph{text strings} recognised by the web application
is given in the first two columns of Table~\ref{tab:ConSQL}.

The formula has to be written in pure text, but thanks to MathJax
the formula will be displayed nicely.

\begin{figure}[h!]
 \centering
 \includegraphics[scale=0.6]{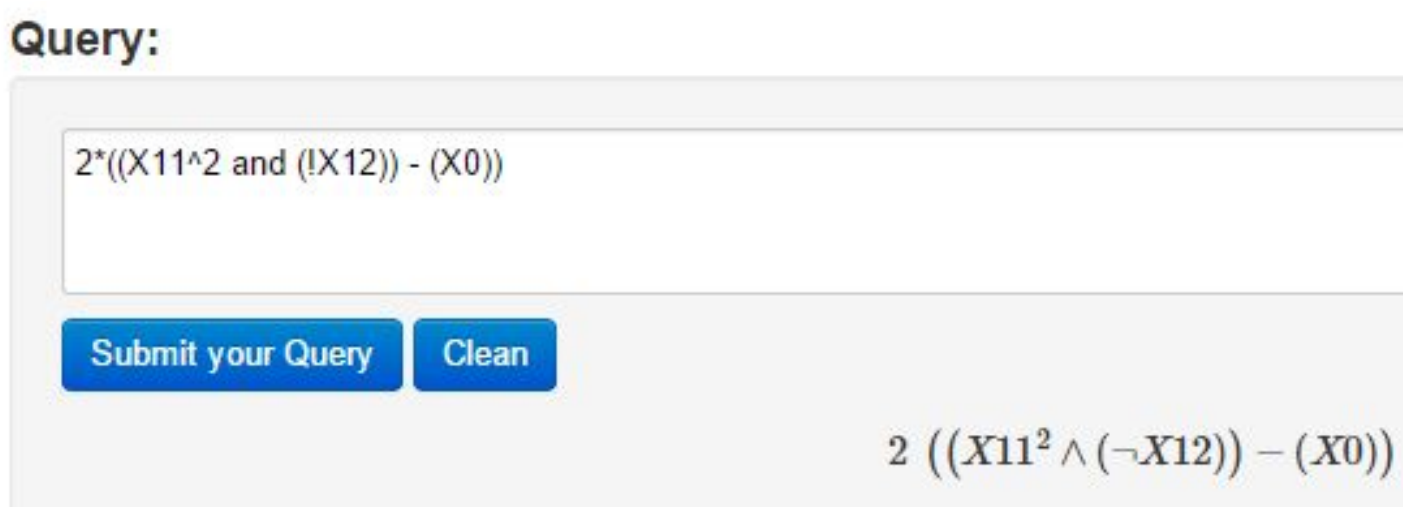}
 \caption{Form for the insertion of queries}
\end{figure}

This formula is then parsed and translated in an SQL query according the association fields-variables
given in Table~\ref{tab:cardata}, and to the following table:

\begin{table}[h!]
\caption{ }\label{tab:ConSQL}
\begin{center}\begin{tabular}{lll}
\hline
Connective  & Text strings		&  SQL statement \\
\hline
$\neg \alpha$   		&\texttt{!}	&\texttt{1 - ($\alpha$)}\\
$\alpha\to \beta$  		&\texttt{->}	&\texttt{least(1,1-($\alpha$- $\beta$))} \\
$\alpha\vee \beta$   		&\texttt{or}	&\texttt{greatest($\alpha$,$\beta$)}\\
$\alpha\wedge \beta$   		&\texttt{and}	&\texttt{least($\alpha$,$\beta$)}\\
$\alpha\leftrightarrow \beta$	&\texttt{<->}	&\texttt{1-ABS($\alpha$-$\beta$)}\\
$\alpha\oplus \beta$  		&\texttt{+}	&\texttt{least(1,$\alpha$+$\beta$)} \\
$\alpha\odot \beta$ 		&\texttt{ox}	&\texttt{greatest(0,$\alpha$+$\beta$-1)}\\
$\alpha\ominus \beta$ 		&\texttt{-}	&\texttt{greatest(0,$\alpha$-$\beta$)}
\end{tabular}\end{center}
\end{table}
\noindent where $\alpha$ and $\beta$ are formulas built from the variables \texttt{X0},\dots,\texttt{X14}.
%
%
For instance, the formula:
\begin{equation*}
\texttt{X1 and (X5 or !X7)}
\end{equation*}
will be translated into the SQL query:
\begin{verbatim}
SELECT id, trim, length, seats, horsepower,
least(length,greatest(seats,horsepower))
As Results FROM auto;
\end{verbatim}

The record set is then displayed as a table in the web page.

\begin{figure}[h!]
 \centering
 \includegraphics[scale=0.38]{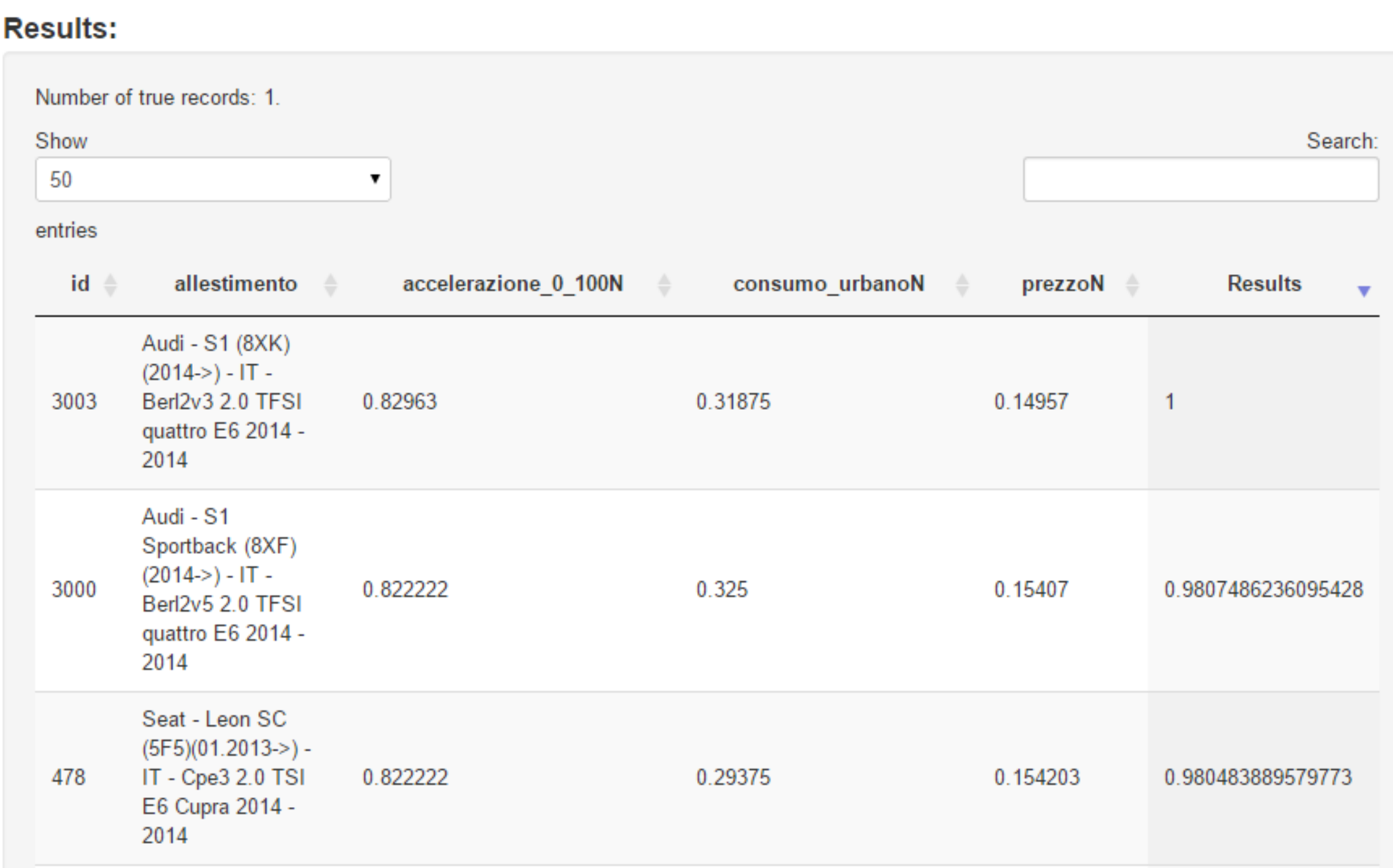}
 \caption{The record set}
\end{figure}

We observe that the iterated conjunction \texttt{$\alpha$\^{}n} is translated
into the SQL expression \texttt{greatest(0,n*$\alpha$-n+1)}. Similarly, the iterated
disjunction \texttt{n$\alpha$} is translated into \texttt{least(1,n*$\alpha$)}. Whence, the translation
of a \L{}ukasiewicz fomula (written using connectives in Table \ref{tab:ConSQL} and the
iterated ones)
into an SQL query takes only linear time.

\section{Case studies}\label{sec:cases}
\subsection{Rewriting a traditional query with {\L}ukasiewicz logic}

Traditionally, an online database of cars is queried via the following steps. First, one
select a (possibly empty) set of filters in order to restrict the search range.
For example, one could choose to look only at diesel powered SUVs. After this preliminary
step, one is asked to apply a conjunction of other filters, expressed in the form of
constraints. For instance, one could ask that the price is between 30,000{\euro} and
50,000{\euro}, \emph{and} that the maximum speed is over 200 km/h. Two major Italian
websites designed along these lines are
Quattroruote\footnote{\texttt{http://www.quattroruote.it/listino/}} and Repubblica
Motori\footnote{\texttt{http://listino-motori.repubblica.it/}}.

Our first basic example aims to describe how to translate a classical query in our
language, \textit{i.e.} in a formula of \L{}ukasiewicz logic. Suppose that our priority is
an excellent acceleration, if possible accompanied with low urban consumption. In our
normalised database, we may, for instance, query with the following instruction
\begin{equation}\label{eq:tradQ}
\texttt{(0.875<=X11) and (X12<=0.25)}
\end{equation}
Under a linear normalisation of the data to the unit interval $[0,1]$ this corresponds with cars whose acceleration 0-100 \emph{km/h} is 5.10 \emph{sec}, or better, and whose urban
consumption is under 8 \emph{l} per 100 \emph{km}. If we query our database with \eqref{eq:tradQ}, the
answer set is
\begin{table}[h!]\begin{tabular}{ll}
2969 &Audi SQ5 (8R) 3.0 TDI DPF quattro 2012 \\
4272 &BMW Serie 3 GT 335d xDrive 2014 \\
4275 &BMW Serie 3 GT 335d xDrive 2014
\end{tabular}\end{table}

The latter two cars are two different trim level of the same model of car.
In this specific case, we can obtain exactly the same result using
a query in \L{}ukasiewicz logic. We first find a query capable of discard all
non-valid cars:
\begin{equation}\label{eq:LukQ_0}
\texttt{X11\^{}8 and (!X12)\^{}4}
\end{equation}
In logical terms, the truth value of this formula when evaluated in a car which does not
belong to our answer set is 0. Then, we need to ensure that the truth value of the formula
is 1 when the formula is evaluated on a car belonging to the answer set. We obtained the
desired result by the query:
\begin{equation}\label{eq:LukQ_1}
\texttt{20(X11\^{}8 and (!X12)\^{}4)}
\end{equation}
In general, is it always possible to rewrite a traditional query in the form
\texttt{X$\leq k$} or \texttt{X$\geq k$} , for $k\in[0,1]$ a rational value, in a \L{}ukasiewicz query which provide
exactly the same result. (For theoretical background on this, please refer to Section \ref{sec:literals}.) Nevertheless, this procedure produces very artificial values for
exponents and multipliers, wiping out any possibility of exploiting in a totally natural
way the expressive potential of the language of \L{}ukasiewicz logic.

In this regard it is worth to observe that, when stepping from the query \eqref{eq:LukQ_0}
to \eqref{eq:LukQ_1}, we lose information. Indeed, the query \eqref{eq:LukQ_0} evaluated
in the answer set is capable to rank the cars. In our case, we obtain that, though
all three car are compatible with our queries, the BMWs better fit our desires.
The answer set of \eqref{eq:LukQ_0}, with the assigned truth valued in square brackets,
is, indeed,
\begin{table}[h!]\begin{tabular}{llc}
4272 &BMW Serie 3 GT 335d xDrive 2014 &[0.170] \\
4275 &BMW Serie 3 GT 335d xDrive 2014 &[0.170]\\
2969 &Audi SQ5 (8R) 3.0 TDI DPF quattro 2012 &[0.052]
 \end{tabular}\end{table}

A further consideration deserves to be done, concerning the query \eqref{eq:LukQ_0}. Such
query is obtained with the exact purpose of rewriting the traditional query
\eqref{eq:tradQ}. But, almost always, our starting point is the expression of
our \emph{desiderata} in natural language: \emph{I want a car with remarkable acceleration, but with reduced urban fuel consumption}. The more natural way to query our database is thus to ask
\begin{equation}\label{eq:LukQ_2}
\texttt{X11\^{}2 and (!X12)}
\end{equation}
The first 10 records of the answer set of \eqref{eq:LukQ_2} are:

\begin{table}[h!]\begin{tabular}{llc}
4272	&BMW Serie 3 GT 335d xDrive 2014	&[0.793]\\
4275	&BMW Serie 3 GT 335d xDrive 2014	&[0.793]\\
2969	&Audi SQ5 (8R) 3.0 TDI DPF quattro 2012 &[0.763]\\
4287	&BMW Serie 6 Coupe 640d xDrive 2012	&[0.748]\\
4369	&BMW Serie 6 GC 640d xDrive 2013	&[0.748]\\
4433	&BMW X4 xDrive 3.5d 2014 &[0.748]\\
4462	&BMW Serie 4 Cabrio 435d xDrive 2014 &[0.748]\\
4468	&BMW Serie 4 Cabrio 435d xDrive 2014 &[0.748]\\
665	&Infiniti Q50 S Hybrid 2013 &[0.737]\\
3028	&Audi A7 Sportback 3.0 TDI quattro 2012 &[0.733]
 \end{tabular}\end{table}

Not surprising, the first three cars in the (ranked) answer set of this query coincides
with the answer set of the traditional query.

Another observation is in order here.
We notice that the query (\ref{eq:LukQ_1}) uses quite complex
linguistic hedges. This may seem highly unnatural.
On the other hand
the numeric-threshold query (\ref{eq:tradQ}) uses a three-digit precision.
One may wonder whether and how this precision is needed, and if it really is,
how it came about.
One can imagine a process of trial and errors, where the trials $\texttt{(0.87<=X11)}$
and $\texttt{(0.88<=X11)}$ fail to produce any usable solution.
Or one can imagine some algorithm that comes to produce this threshold via some
sophisticated heuristics able to sift through to the third decimal before even submitting the query.
In any case, whatever the process, it seems that if a precision of many decimals is really needed
then the determination of the query requires a non-negligible additional computational effort,
either performed by the human user, or by the machine.
We do not want to embark in the task of measuring precisely this effort,
but we only offer the remark that the growing complexity we see in the use of many decimals
is reflected in the growing complexity of the linguistic hedges used in the {\L}ukasiewicz query.
One query should be deemed as natural as the other one.
The fact that we have an infinite reservoir of distinct hedges then comes
as a useful tool if we ever have to gauge complexity of threshold determination.

\subsection{Exploiting the power of {\L}ukasiewicz logic}

Consider the query \eqref{eq:LukQ_2}. Suppose we want to add more conditions to our request.
Namely, suppose we do not want to spend so much money for our car, but we are looking, instead, for something \emph{very cheap}. We can model our request by adding to \eqref{eq:LukQ_2}
a new clause talking about the cost of the car, as follows.
\begin{equation}\label{eq:LukQ_3}
\texttt{X11\^{}2 and (!X12) and (!X0)\^{}3}
\end{equation}
The answer set seems to satisfy our request (the first 10 results are displayed):

\begin{table}[h!]\begin{tabular}{llc}
4688	&Ford Focus ST EcoBoost 250Cv 2012 &[0.556]\\
3003	&Audi S1 2.0 TFSI quattro 2014 &[0.551]\\
2745    &Volkswagen Golf VII 2.0 TSI  GTI 2013 &[0.546]
 \end{tabular}
\end{table}
\begin{table}[h!]\begin{tabular}{llc}
2205	&Renault Megane III Coupe 2.0 TCe 265 2014 &[0.544]\\
4781	&Ford Focus Wagon ST 2012 &[0.541]\\
3000	&Audi S1 Sportback 2.0 TFSI quattro 2014 &[0.538]\\
478	&Seat Leon SC TSI Cupra 2014	&[0.537]\\
3292	&Opel Astra J GTC Turbo OPC 2012	&[0.535]\\
496	&Seat Leon 2.0 TSI Cupra 2014 [&0.530]\\
2736	&Volkswagen Golf VII 2.0 TSI  GTI 2013 &[0.528]\\
52	&Renault Clio IV 1.6 TCe 200 Monaco GP 2014 &[0.526]
 \end{tabular}
\end{table}

If we want, in addition, the cubic capacity to be \emph{very small}, we can use the query
\begin{equation}\label{eq:LukQ_4}
\texttt{X11\^{}2\;and\;(!X12)\;and\;(!X0)\^{}3\;and\;(!X6)\^{}2}
\end{equation}
obtaining (the first 6 results are displayed):

\begin{table}[h!]\begin{tabular}{llc}
52	&Renault Clio IV 1.6 TCe 200 Monaco GP 2014 &[0.526] \\
54	&Renault Clio IV 1.6 TCe 200 R.S. 2013 &[0.526] \\
2395	&Renault Clio IV 1.6 TCe 200 R.S. 2013 &[0.526] \\
3967	&MINI CLUBMAN 2010 &[0.511] \\
1196	&MINI COUPE JCW 2011	&[0.507] \\
448	&Seat Ibiza SC 1.4 TSI SC Cupra 2013	&[0.496]
 \end{tabular}\end{table}

\medskip
After these simple examples, two main consideration are in order.
First, queries written as formul\ae{} in \L{}ukasiewicz logic are easily updatable.
In a traditional query like \eqref{eq:LukQ_0}, based on the enforcement of crisp thresholds
to the designated features, one usually have to reformulate the thresholds before
stating some new conditions. This is especially true in our specific case,
where the answer set
contains as few as 3 cars, and there is no chance to regain new answers where asking for
cars that are \emph{also} cheap.
The adaptive behavior of our queries is instead very natural. Of course, when we conjunct
new conditions, the truth values of the top records of the answer set is decreasing. This
is quite intuitive, since the more we are demanding, the less there is the possibility
for our query to be fully satisfied by a car. We shall overcome this obstacle by asking
for our final query to be \emph{somewhat satisfied}. To do so, if $\alpha$ is the query,
we simple need to write $n * \alpha$, with $n>1$: the bigger is $n$, the more we relax
our query. With this technique we can also get a query that is fully satisfied
(\emph{i.e.}, in degree $1$) by a non-empty set of cars. For instance, the query
\begin{equation*}
\texttt{2*(X11\^{}2\,and\,(!X12)\,and\,(!X0)\^{}3\,and\,(!X6)\^{}2)}
\end{equation*}
returns the following

\begin{table}[h!]\begin{tabular}{llc}
52	&Renault Clio IV 1.6 TCe 200 Monaco GP 2014 &[1]\\
54	&Renault Clio IV 1.6 TCe 200 R.S. 2013 &[1]\\
2395	&Renault Clio IV 1.6 TCe 200 R.S. 2013 &[1]\\
3967	&MINI CLUBMAN 2010 &[1]\\
1196	&MINI COUPE JCW 2011	&[1]\\
448	&Seat Ibiza SC 1.4 TSI SC Cupra 2013	&[0.993]
 \end{tabular}
\end{table}

\medskip
The second consideration: it is easy and natural dealing with linguistic
hedges. A linguistic hedge is a mitigating or intensifying word used to lessen or
increase the impact of a sentence (see \cite{hedge} for a fuzzy-set-theoretic
interpretation of linguistic hedges).
As described in the introduction, \L{}ukasiewicz logic has the expressive power to
deal with expressions like \emph{very}, or \emph{somewhat}. Consider the
query \eqref{eq:LukQ_4}. There, we are asking for a car with a
\emph{very} good acceleration, with low urban consumption, which is \emph{very very}
cheap, and with a \emph{very} small cubic capacity. Suppose we want to lighten some
requests. Suppose, for instance, that we can fit with a car with a
good acceleration, which is \emph{very} cheap, with a small cubic capacity and with urban consumption that are \emph{somewhat} low. We query with
\begin{equation}\label{eq:LukQ_5}
\texttt{X11\;and\;2*(!X12)\;and\;(!X0)\^{}2\;and\;(!X6)}
\end{equation}
and obtain (the first 6 results are displayed):

\begin{table}[h!]\begin{tabular}{llc}
52	&Renault Clio IV 1.6 TCe 200 Monaco GP 2014 &[0.763]\\
54&Renault Clio IV 1.6 TCe 200 R.S. 2013 &[0.763]\\
2395	&Renault Clio IV 1.6 TCe 200 R.S. 2013 &[0.763]\\
448	&Seat Ibiza SC 1.4 TSI SC Cupra 2013	&[0.748]\\
4720	&Ford Fiesta 1.6 EcoBoost ST 2013 &[0.748]\\
1198	&MINI COUPE S 2011 &[0.744]
 \end{tabular}
\end{table}

As expected, results are similar, though with higher values of truth degrees.

\subsection{Towards a better understanding of the intended semantics of {\L}ukasiewicz logic}

We discuss here the r\^ole of the \emph{minus} connective $\ominus$. Our starting point is the query
\eqref{eq:LukQ_2}. Let us state that \eqref{eq:LukQ_2} describe our notion of
a \emph{good} car. That is, a car is \emph{good} if fully meets our demand, it is
\emph{poor} if it does not satisfy our request at all, it is \emph{medium},
otherwise. Like in the preceding paragraph, we want to add a further request: that
the car is economic. However, we aim to state something different from
what is stated by
the query \eqref{eq:LukQ_3}. We do not want, in fact, to  force the car to be cheap.
We want, instead, to assert that we do not make any difference between a good car
with a medium price, and a medium car with a low price. To make this request, we
use the following query.
\begin{equation}\label{eq:LukQ_6}
\texttt{(X11\^{}2 and (!X12)) - (X0)}
\end{equation}
According to the semantics presented in the introduction (and remembering that
the left part of the expression represent our notion of a good car), we are asking
for a car that is much more good than expensive.
Thus, the perfect matching is given by
a \emph{top} car (as before, according to our notion) with a ridiculously low
price. Most probably, such a car does not exist. As a matter of fact,
the answer we obtain querying the database is the following.

\begin{table}[h!]\begin{tabular}{llc}
3003	&Audi S1 2.0 TFSI quattro 2014 &[0.510]\\
3000	&Audi S1 Sportback 2.0 TFSI quattro 2014 &[0.490]\\
478	&Seat Leon SC 2.0 TSI Cupra 2014	&[0.490]\\
496	&Seat Leon 2.0 TSI Cupra 2014 &[0.488]\\
2988	&Audi S3 2.0 TFSI quattro 2013	&[0.481]\\
487	&Seat Leon SC 2.0 TSI Cupra 2014 &[0.480]\\
4275	&BMW Serie 3 GT 335d xDrive 2014	&[0.480]\\
497	&Seat Leon 2.0 TSI Cupra 2014 &[0.478]\\
2993	&Audi S3 Sportback 2.0 TFSI quattro 2013 &[0.476]\\
1595	&BMW Serie 2 Coupe 228i 2014 &[0.472]
 \end{tabular}\end{table}

None of the cars in our database fully fit the request. Though, as
mention in the previous subsection, we could easily lighten the request to
\begin{equation*}
\texttt{2*((X11\^{}2 and (!X12)) - (X0))}
\end{equation*}
obtaining, as the only car satisfying (at $1$) our formula:

\medskip
\begin{table}[h!]\begin{tabular}{llc}
3003  &	Audi S1 2.0 TFSI quattro 2014 &[1]
 \end{tabular}\end{table}

We want to stress the difference between \eqref{eq:LukQ_6} and the following query,
seemingly similar in meaning.
\begin{equation}\label{eq:LukQ_7}
\texttt{X11\^{}2 and (!X12) and (!X0)}
\end{equation}
The top ranked cars according to the latter query are

\begin{table}[h!]\begin{tabular}{llc}
4270	&BMW Serie 3 GT 330d xDrive 2013 &[0.711]\\
3685&Mercedes-Benz A 45 AMG 2012 &[0.700]\\
665&	Infiniti Q50 S Hybrid 2013 &[0.698]
 \end{tabular}\end{table}

while 3003, the only car satisfying \eqref{eq:LukQ_6}, ends up here in 20$^{th}$ position.
Comparing those entries we discover that while 4270 is better than 3003, the
latter is distinguished by the following property: it maximises the difference between the truth
value of the formula stating (in short) that \emph{the car is good}, and the formula asserting that
\emph{the car is expensive}.

A more general example may clarify this point. Suppose you can describe, via a formula $\alpha$,
what (in your opinion) makes a car a good car. Further, suppose you can describe, via
a formula $\beta$, what makes a car a bad car. Consider the formula $\gamma = \alpha \wedge \neg \beta$.
Intuitively, $\gamma$ describe your perfect car, with all pros and no cons. Let $A$ be a car satisfying $\alpha$
in degree $0.5$, and $\beta$ in degree $0.5$. Thus, $A$ satisfy $\gamma$ in degree $0.5$.

Let now $\varphi=\alpha - \beta$. Remember that, in natural language, $\varphi$ asks for a car which
is \emph{much more $\alpha$ than $\beta$}. Observe that the truth value of $\varphi$ when evaluated in $A$ is $0$.
While according to $\gamma$, the car $A$ is a medium car, $\varphi$ discard the same car. While $\gamma$
simply measures the degree in which the car fits our collection (conjunction) of requests, the formula
$\varphi$ add a higher level to the query measuring the distance between
the two simpler queries $\alpha$ and $\beta$. An assertion like $\varphi$ can be, in fact,
interpreted as a measure of the level of satisfaction of the buyer. In essence, the buyer get no satisfaction
in buying whichever car with equivalent pros and cons. Satisfaction begin when  pros overcome cons,
and is maximal when we have all possible pros, and no cons. When dealing with queries constructed in the manner
of $\varphi$, this is essentially how the logical connective $\ominus$ may serve us.

To stress this point, we provide another example.
Assume the buyer asks for a car which is both $\gamma$ and $\delta$, say, "a powerful but economic car"
(that is, "a powerful but not expensive car").
On further thinking the buyer makes the additional point that "in any case I prefer the car being economic
rather than powerful", that can be modeled by $\gamma \ominus \neg\delta$ (or, equivalently,
by $\delta \ominus \neg\gamma$), possibly attenuated by the use of a {\em somewhat} hedge.
With this structure it appears clearly that use of $\ominus$ provides, semantically, a higher
level in the language, since the clarification added by the buyer directly refers to the
evaluation and weighing of the two simpler queries $\gamma$ and $\delta$.
This cannot be achieved with the use of the other connectives $\neg,\vee,\wedge$ only.


\balance
\section{Conclusion}
The goal of this work is twofold. The case studies we proposed provide a better
understanding of the intended semantics presented in the introduction.
Indeed, in appreciating the answer given by our database system to our queries,
we make clearer the natural linguistic meaning of a formula of \L{}ukasiewicz
logic. Moreover, we provide a further, pragmatic, justification of the proposed semantics.

With this work, we also aim to convince the reader that the pure
\L{}ukasiewicz logic can be effectively, and efficiently, used to treat real
problems: in our case, to query an online database of cars. (Other, distinct, cases
where \L{}ukasiewicz logic is used in applications can be found in literature.
See, for instance,
\cite{c_bcmm_fuzz_2010a,c_bcmm_fuzz_2010b}, where the use of \L{}ukasiewicz implication in
control systems is first theoretically discussed and then analyzed in an experiment on the
technical analysis of the financial markets. Another example is \cite{turunen},
where a \L{}ukasiewicz many-valued logic similarity based fuzzy control algorithm
is introduced, and tested in three realistic traffic signal control systems.
See also \cite{godo,russo}.)
Much work remains to be done on this front to develop an online system which is meant for a general
user, typically not enough expert in logic to query with logical formul\ae.
First, we need to design the mechanisms of interaction between user and system (the
way in which the user will form its query). This will probably be done with repeated
questions to the user, in order to identify the most of her/his desires, her/his
preferences, and the issues to which it is indifferent. Following this, the system
must be equipped with an algorithm able to translate the user's desiderata in
a \L{}ukasiewicz formula. Only at this point, the system we developed and presented
in this
paper will go into action, providing an answer to the query.

\section*{Acknowledgment}
We thank Vincenzo Marra for many useful discussions on the topics of this paper.

\medskip
The authors were supported by the MIUR-FIRB research project \emph{PNCE - Probability theory of non-classical events}.
Valota acknowledges also partial support from a Marie Curie INdAM-COFUND Outgoing Fellowship. The research reported in this paper was carried out while Valota was a postdoc fellow of the Dipartimento di Scienze Teoriche e Applicate (Universit\`{a} dell'Insubria), supported by the PNCE research project.

\bibliographystyle{IEEEtran}
\bibliography{fuzzdb}

\end{document}